\documentclass[11pt]{scrartcl}  
\usepackage[left=1in, right=1in, top=1in, bottom=1in]{geometry}
\usepackage{mathptmx} 
\usepackage{multirow,graphicx,physics,hyperref,amsmath,amssymb,graphicx,mathalfa,mathtools,colortbl,hhline,bbold}
\numberwithin{equation}{section}
\usepackage{hyperref}
\hypersetup{colorlinks=true, citecolor=blue}
\usepackage{array}

\usepackage{longtable,caption}
\newcolumntype{L}[1]{>{\raggedright\let\newline\\\arraybackslash\hspace{0pt}}m{#1}}
\newcolumntype{C}[1]{>{\centering\let\newline\\\arraybackslash\hspace{0pt}}m{#1}}
\newcolumntype{R}[1]{>{\raggedleft\let\newline\\\arraybackslash\hspace{0pt}}m{#1}}
\usepackage[framemethod=TikZ]{mdframed}
\mdfdefinestyle{sid}{%
    linecolor=black,
    outerlinewidth=1.0pt,
    roundcorner=7pt,
    innerrightmargin=15pt,
    innerleftmargin=15pt,
    backgroundcolor=gray!30
		}	
\definecolor{Gray}{gray}{0.8}
\definecolor{MyBlue}{rgb}{0.0,0.0,0.9}
\definecolor{MyRed}{rgb}{0.0,0.9,0.0}
\colorlet{Bluee}{MyBlue!6}
\colorlet{Redd}{MyRed!1}
\DeclareMathAlphabet{\mathcal}{OMS}{cmsy}{m}{n} 

\usepackage{xcolor}
\colorlet{sectioncolor}{blue!20}
\colorlet{subsectioncolor}{orange!70}
\colorlet{subsubsectioncolor}{green!40}
\makeatletter
\renewcommand\sectionlinesformat[4]{%
  \colorbox{#1color}{%
    \parbox[t]{\dimexpr\textwidth-2\fboxsep\relax}{%
      \raggedsection\color{black}\@hangfrom{#3}{#4}%
}}}
\makeatother
\usepackage{blindtext}
\title{\textbf{Analyzing reduced density matrices in SU(2) Chern-Simons theory}}
\date{}
\author{\textbf{\emph{Atesh Saini}} and \textbf{\emph{Siddharth Dwivedi}}\thanks{Email: siddharth.dwivedi@curaj.ac.in, ateshsaini6@gmail.com}\\\\ Department of Physics, School of Physical Sciences, \\ Central University of Rajasthan, Ajmer, 305817, India}
\begin{document}
\maketitle
\begin{abstract}
We investigate the reduced density matrices obtained for the quantum states in the context of 3d Chern-Simons theory with gauge group SU(2) and Chern-Simons level $k$. We focus on the quantum states associated with the $T_{p,p}$ torus link complements, which is a $p$-party pure quantum state. The reduced density matrices are obtained by taking the $(1|p-1)$ bi-partition of the total system. We show that the characteristic polynomials of these reduced density matrices are monic polynomials with rational coefficients. 
\end{abstract}
\hypersetup{linkcolor=red}
\listoftables

\hypersetup{linkcolor=blue}
\tableofcontents
\newpage
\section{Introduction} \label{sec1}
The 2+1 dimensional Chern-Simons theory with gauge group $G$ and level $k \in \mathbb{Z}$ is defined on a 3-manifold $M$ with action given by \cite{Witten:1988hf}
\begin{equation}
S(A) = \frac{k}{4\pi} \int_M \text{Tr}\left(A \wedge dA + \frac{2}{3} A \wedge A \wedge A \right) ~,
\label{CSaction}
\end{equation}
where $A = A_{\mu}dx^{\mu}$ is a gauge field, which, in this case, is a connection on the trivial $G$-bundle over $M$. The gauge invariant operators in the theory are  Wilson lines. Given an oriented knot $\mathcal{K}$ embedded in $M$, the Wilson line is defined by taking the trace of the holonomy of $A$ around $\mathcal{K}$ :
\begin{equation}
W_{R}(\mathcal{K}) = \text{Tr}_R \,P\exp\left(i\oint_{\mathcal{K}} A \right) ~,
\end{equation} 
where the trace is taken under the representation $R$ of $G$. The computation of the partition function in Chern-Simons theory involves the integration over the infinite-dimensional space of connections. For the bare manifold $M$ without any Wilson line, the partition function is given as
\begin{equation}
Z(M) = \int e^{i S(A)} dA ~,
\end{equation}
where $dA$ is an appropriate quantized measure defined for a connection $A$, and the integration is over all the gauge invariant classes of connections. Since the connection has been integrated out, $Z(M)$ is a topological invariant of $M$. One can also compute the partition function of $M$ in the presence of knots and links by inserting the appropriate Wilson loop operators in the integral. Given a link $\mathcal{L}$ made of disjoint oriented knot components, i.e. $\mathcal{L} = \mathcal{K}_1 \sqcup \mathcal{K}_2 \sqcup \ldots \sqcup \mathcal{K}_n$, the partition function of $M$ in the presence of $\mathcal{L}$ can be obtained by modifying the path integral as:
\begin{equation}
Z(M; \mathcal{L}) = \int e^{i S(A)}\, W_{R_1}(\mathcal{K}_1)\ldots W_{R_n}(\mathcal{K}_n) \, dA ~.
\end{equation}  
In this article, we are interested in manifolds $M$ with boundary $\Sigma$. For this, we must incorporate the boundary condition $A\rvert_{\Sigma} = Q$ imposed on $\Sigma$ while performing the path integral. In such a case, the partition function on $M$ with the Wilson line insertions is interpreted as the wave function of a state. Hence, we write the quantum state as:
\begin{equation}
\ket{\Psi} \equiv Z_Q(M; \mathcal{L}) = \int_{A\rvert_{\Sigma} = Q} e^{i S(A)}\, W_{R_1}(\mathcal{K}_1)\ldots W_{R_n}(\mathcal{K}_n) \, dA ~.
\end{equation}
This state is an element of the Hilbert space $\mathcal{H}_{\Sigma}$ associated with $\Sigma$. When the boundary is a torus, i.e., $\Sigma=T^2$, the basis of the Hilbert space can be defined by doing the Chern-Simons path integral on a solid torus with Wilson lines inserted along its contractible cycle. These Wilson lines carry certain allowed representations of the gauge group $G$ for a given Chern-Simons level $k$. These are called `integrable representations'. For the SU(2) group and level $k$, the integrable representations are given by the spin $a/2$ where $a=0,1,2,\ldots,k$. The path integral on solid torus with the Wilson line inserted along its contractible cycle and carrying spin $a/2$ representation will give the basis state $\ket{e_{a}}$ of the Hilbert space $\mathcal{H}_{T^2}$. Thus the Hilbert space is of dimension $(k+1)$ with basis given as:
\begin{equation}
\text{basis}(\mathcal{H}_{T^2}) = \{\ket{e_0}, \ket{e_1}, \ket{e_2}, \ldots, \ket{e_k} \}~.
\end{equation}
Note that there can be many topologically distinct manifolds with a torus boundary. Each such manifold will give a quantum state in $\mathcal{H}_{T^2}$. Hence, given the manifold $M$, we will denote the corresponding state as $\ket{M} \in \mathcal{H}_{T^2}$. 

When $M$ is a knot complement $S^3 \backslash \mathcal{K}$, the corresponding state $\ket{S^3 \backslash \mathcal{K}}$ can be expanded as \cite{Balasubramanian:2016sro}:
\begin{equation}
\ket{S^3 \backslash \mathcal{K}} = \sum_{a=0}^k J_a(\mathcal{K})\, \ket{e_a} ~,
\label{State-Knot}
\end{equation}
where $J_a(\mathcal{K})$ is the colored Jones invariant of the knot $\mathcal{K}$, carrying the spin $a/2$ representation of SU(2). Note that the colored Jones polynomials depend on the framing of the knot. These factors (though representation dependent) act as a unitary transformation on the basis of the Hilbert space and will not affect the entanglement entropy as illustrated in \cite{Balasubramanian:2016sro}. Hence, we can neglect the framing factors or overall factors in \eqref{State-Knot}. Moreover, since we are performing the computations on the Chern-Simons theory side, the variable $q$ in the colored Jones polynomial must be set to $q=\exp(\frac{2\pi i}{k+2})$ as discussed in \cite{Witten:1988hf}. There is a straightforward generalization of \eqref{State-Knot} to link complements. Let $\mathcal{L}$ be an $n$-component link  $\mathcal{L} = \mathcal{K}_1 \sqcup \mathcal{K}_2 \sqcup \ldots \sqcup \mathcal{K}_n$. Then, the quantum state $\ket{S^3 \backslash \mathcal{L}}$ associated with the link complement will be an element of the tensor product of $n$ copies of $\mathcal{H}_{T^2}$ and will be expanded as:
\begin{equation}
\ket{S^3 \backslash \mathcal{L}} = \sum_{a_1=0}^k\sum_{a_2=0}^k\cdots\sum_{a_n=0}^k J_{a_1,a_2,\ldots,a_n}(\mathcal{L})\, \ket{e_{a_1},e_{a_2},\ldots,e_{a_n}} ~,
\label{State-Generic}
\end{equation}
where $J_{a_1,a_2,\ldots,a_n}(\mathcal{L})$ is the colored Jones polynomial of link $\mathcal{L}$ whose individual knots carry SU(2) spin representations $a_1/2, a_2/2, \ldots, a_n/k$ respectively. Again, the framing factors of individual knots do not affect the entanglement measures. The notation $\ket{e_{a_1},e_{a_2},\ldots,e_{a_n}}$ is a shorthand notation for $\ket{e_{a_1}}\otimes \ket{e_{a_2}} \otimes \ldots \otimes \ket{e_{a_n}}$.

The setup discussed above provides a tractable computational setup to calculate the quantum state, obtain the reduced density matrix (RDM) by tracing out a subset of Hilbert spaces, and study the entanglement properties. For the advances in this field of study, we refer the readers to \cite{Balasubramanian:2016sro,Dwivedi:2017rnj,Balasubramanian:2018por,Hung:2018rhg,Melnikov:2018zfn,Camilo:2019bbl,Dwivedi:2019bzh,Buican:2019evc,Zhou:2019ezk,Dwivedi:2020jyx,Dwivedi:2020rlo,Dwivedi:2021dix,Dwivedi:2024gzg}. This line of study might be useful in giving an insight to the more general open problem of classification of entanglement in quantum field theories. 

Apart from the connection with knot theory, the topological entanglement entropies computed in this setup have interesting number-theoretic properties, which can be seen in \cite{Dwivedi:2020rlo} and \cite{Dwivedi:2024gzg}. Note that although \eqref{State-Generic} is straightforward, the calculations are not. This is because to obtain the state $\ket{S^3\backslash \mathcal{L}}$ itself, we must know the colored Jones polynomials of link $\mathcal{L}$ for arbitrary representations on various knot components. This data is limited, and as the value of $k$ increases, obtaining the colored Jones polynomials for higher colors becomes more and more computationally involved. To circumvent this difficulty, we would like to consider a link whose colored Jones is readily available for all colors. The torus links provide an elegant class of such type of links whose colored Jones invariants can be explicitly computed from \cite{Stevan:2010jh,Brini:2011wi}.

In the present article, we will focus on torus links of type $T_{p,p}$. The torus link $T_{p,p}$ consists of $p$ number of circles such that the linking number between any two circles is 1. The link complement $S^3 \backslash T_{p,p}$ will have $p$ number of disjoint torus boundaries. Therefore, the associated quantum state $\ket{S^3 \backslash T_{p,p}}$ lives in the tensor product of $p$ copies of $\mathcal{H}_{T^2}$. Tracing out a subset of Hilbert space will give us a reduced density matrix $Y_p$. In this work, we plan to investigate more about the matrices $Y_p$. Our results and findings are given in section \ref{sec2}. Conclusion and open problems are discussed in section \ref{sec3}.
\section{Characteristic polynomials of RDM for torus links $T_{p,p}$ } \label{sec2}
As discussed in \eqref{State-Generic}, the quantum state associated with the manifold $S^3 \backslash T_{p,p}$ will be given as:
\begin{equation}
\ket{S^3 \backslash T_{p,p}} = \sum_{a_1=0}^k\sum_{a_2=0}^k\cdots\sum_{a_p=0}^k J_{a_1,a_2,\ldots,a_p}(T_{p,p})\, \ket{e_{a_1},e_{a_2},\ldots,e_{a_p}} ~.
\label{State-Tpp}
\end{equation}
The colored Jones invariant of the torus link $T_{p,p}$ is given as \cite{Stevan:2010jh,Brini:2011wi}:
\begin{equation}
 J_{a_1,a_2,\ldots,a_p}(T_{p,p}) = \sum_{b=0}^k \sum_{c=0}^k \frac{\mathcal{S}_{0b}\, \mathcal{T}_{bb}\, \mathcal{S}_{bc}}{(\mathcal{S}_{0c})^{p-1}} \left(\mathcal{S}_{c\,a_1} \, \mathcal{S}_{c\,a_2} \, \ldots \, \mathcal{S}_{c\,a_p} \right) ~.
 \label{Jones-Tpp}
\end{equation}
Here $\mathcal{S}_{\alpha \beta}$ and $\mathcal{T}_{\alpha \beta}$ denote the matrix elements of the generators $\hat{\mathcal{S}}$ and $\hat{\mathcal{T}}$ of the unitary representation of the modular group SL(2, $\mathbb{Z}$). These generators act as diffeomorphism operators for the torus $T^2$, and their matrix elements for SU(2) gauge group and Chern-Simons level $k$ are given as \cite{Dwivedi:2017rnj}:
\begin{align}
\mathcal{S}_{\alpha \beta} &= \sqrt\frac{2}{k+2}\, \sin\left[\frac{\pi(\alpha+1)(\beta+1)}{k+2} \right] \nonumber \\ 
\mathcal{T}_{\alpha \beta} &= \exp\left[\frac{i \pi  \alpha (\alpha+2)}{2 (k+2)} -\frac{i \pi  k}{4 (k+2)} \right] \delta_{\alpha \beta}   ~.
\end{align}
Here $\alpha$ and $\beta$ denotes the spin $\alpha/2$ and spin $\beta/2$ representations of SU(2). Note that as discussed before, we do not care about any framing factors or overall factors present in the invariants \eqref{Jones-Tpp} as they will not affect the spectrum of reduced density matrices. We can simplify the invariants by writing them as follows:
\begin{equation}
 J_{a_1,a_2,\ldots,a_p}(T_{p,p}) = \sum_{c=0}^k \frac{(\mathcal{S}\mathcal{T}\mathcal{S})_{0c}}{(\mathcal{S}_{0c})^{p-1}} \left(\mathcal{S}_{c\,a_1} \, \mathcal{S}_{c\,a_2} \, \ldots \, \mathcal{S}_{c\,a_p} \right) ~.
\end{equation}
We also know that $\mathcal{S}$ and $\mathcal{T}$ are unitary and symmetric matrices. Additionally, for SU(2) group, they satisfy  $\mathcal{S}^2 = (\mathcal{S} \mathcal{T})^3 = \mathbb{1}$, with $\mathbb{1}$ denoting the identity matrix. Using this fact, we can write $\mathcal{S}\mathcal{T}\mathcal{S} = \mathcal{T}^{-1}\mathcal{S}\mathcal{T}^{-1}$. As a result, we will have:
\begin{equation}
 J_{a_1,a_2,\ldots,a_p}(T_{p,p}) = \sum_{c=0}^k \frac{(\mathcal{S}_{0c})^{2-p}}{\mathcal{T}_{00} \mathcal{T}_{cc}} \left(\mathcal{S}_{c\,a_1} \, \mathcal{S}_{c\,a_2} \, \ldots \, \mathcal{S}_{c\,a_p} \right) ~.
 \label{Jones-Tpp-Final}
\end{equation}
For $p=2$, we can show that the above invariant reduces to:
\begin{equation}
 J_{\alpha,\beta}(T_{2,2}) = \sum_{c=0}^k \frac{\mathcal{S}_{\alpha c}\, \mathcal{S}_{c \beta}}{\mathcal{T}_{00}\, \mathcal{T}_{cc}} = \frac{(\mathcal{S}\mathcal{T}^{-1}\mathcal{S})_{\alpha \beta}} {\mathcal{T}_{00}} =  \frac{\mathcal{T}_{\alpha \alpha} \mathcal{T}_{\beta \beta}}{\mathcal{T}_{00}} \mathcal{S}_{\alpha \beta} ~.
\end{equation}
We can recognize this as the Hopf link invariant with framing factors $\mathcal{T}_{\alpha \alpha}$ and $\mathcal{T}_{\beta \beta}$ of the two circles if we ignore the constant factor $\mathcal{T}_{00}$.

Let us now come back to our state \eqref{State-Tpp}. Using \eqref{Jones-Tpp-Final}, we can write it as:
\begin{equation}
\ket{S^3 \backslash T_{p,p}} = \sum_{a_1=0}^k\cdots\sum_{a_p=0}^k \sum_{c=0}^k \frac{(\mathcal{S}_{0c})^{2-p}}{\mathcal{T}_{00} \mathcal{T}_{cc}} \left(\mathcal{S}_{c\,a_1} \, \ldots \, \mathcal{S}_{c\,a_p} \right)\, \ket{e_{a_1},\ldots,e_{a_p}} ~.
\end{equation}
Although this is the final form of the state that we desire, we know that the spectrum of the density matrices does not depend upon a local unitary change of the basis of the Hilbert space. So, we invoke the following unitary transformations on the basis of each of the $p$ Hilbert spaces:
\begin{align}
\ket{e_{a_1}} = \sum_{b_1 = 0}^k \mathcal{S}_{a_1 b_1} \ket{f_{b_1}} \quad,\quad \ldots \quad,\quad \ket{e_{a_p}} = \sum_{b_p = 0}^k \mathcal{S}_{a_p b_p} \ket{f_{b_p}} 
\end{align}
In the new basis, we can expand the quantum state as:
\begin{equation}
\ket{S^3 \backslash T_{p,p}} = \sum_{a_1,\,\ldots,\,a_p=0}^k \sum_{b_1,\,\ldots,\,b_p=0}^k \sum_{c=0}^k \frac{(\mathcal{S}_{0c})^{2-p}}{\mathcal{T}_{00} \mathcal{T}_{cc}} \prod_{i=1}^p\left(\mathcal{S}_{c\,a_i} \,\mathcal{S}_{a_i\,b_i} \right)\, \ket{f_{b_1},\ldots,f_{b_p}} ~.
\end{equation}
Using the fact that $\mathcal{S}^2 = \mathbb{1}$, the sum over variables $a_i$ can be performed. This gives delta functions, which remove the sum over variables $b_i$. Ultimately, we are left with:
\begin{equation}
\ket{S^3 \backslash T_{p,p}} = \sum_{c=0}^k \frac{(\mathcal{S}_{0c})^{2-p}}{\mathcal{T}_{00} \mathcal{T}_{cc}} \, \ket{f_c,f_c,\ldots,f_c} ~.
\end{equation}
Note that in the new basis, the wave function has changed. However, the unitary transformed basis ensures that the spectrum of density matrices and, hence, any entanglement measures will remain unchanged in the new basis.

The first step after getting the quantum state is to normalize it. For this, we must divide it by $\bra{S^3 \backslash T_{p,p}}\ket{S^3 \backslash T_{p,p}}^{1/2}$ which we define as $N_p^{1/2}$. A little calculation shows that
\begin{equation}
N_p  = \sum_{c=0}^k (\mathcal{S}_{0c})^{4-2p} ~.
\end{equation} 
In order to study the entanglement, we need at least two Hilbert spaces, so we must have $p\geq 2$. The total density matrix obtained as $\ket{S^3 \backslash T_{p,p}}\bra{S^3 \backslash T_{p,p}}$ will be a diagonal matrix of order $(k+1)^p$. To obtain the reduced density matrix, we bi-partition the total Hilbert space into two parts  $\mathcal{H}_A$ and $\mathcal{H}_B$ where $\mathcal{H}_A$ is a tensor product of $p_1$ Hilbert spaces and $\mathcal{H}_B$ is the tensor product of remaining $(p-p_1)$ Hilbert spaces. A little exercise shows that the reduced density matrix, which is of order $(k+1)^{p_1}$, is a diagonal matrix that has only $(k+1)$ number of non-zero eigenvalues given as:
\begin{equation}
\lambda_{p,a} = \frac{(\mathcal{S}_{0a})^{4-2p}}{N_p} \quad,\quad a=0,1,2,\ldots,k ~.
\label{eigen-Tpp}
\end{equation}
Thus, the entanglement measures do not depend on the value of $p_1$, and without loss of generality, we can assume the bi-partition $(1|p-1)$ of the total Hilbert space. We call the corresponding reduced density matrix as $Y_p$ whose eigenvalues are given in \eqref{eigen-Tpp}.

Next, we would like to analyze the characteristic polynomial of the density matrices $Y_p$. This polynomial, in variable $x$, is defined as:
\begin{equation}
\text{CP}_p(x) = (x-\lambda_{p,0})(x-\lambda_{p,1})\ldots(x-\lambda_{p,k}) = \prod_{a=0}^k(x-\lambda_{p,a}) ~.
\end{equation}
This polynomial will be a polynomial in variable $x$ of degree $(k+1)$ written as:
\begin{equation}
\text{CP}_p(x) = \sum_{n=0}^{k+1} A_{p,n}\,x^n = A_{p,0}+A_{p,1}\,x+A_{p,2}\,x^2+\ldots+A_{p,k+1}\, x^{k+1} ~,
\end{equation}
where $A_{p,n}$ are the polynomial coefficients and they implicitly depend on the value of $k$. The characteristic polynomial can also be expanded in terms of the `elementary symmetric functions' of the eigenvalues as follows:
\begin{equation}
\text{CP}_p(x) = e_0\,x^{k+1} - e_1\, x^k + e_2 \, x^{k-1} + \ldots + (-1)^{k} e_{k}\,x + (-1)^{k+1} e_{k+1} ~,
\end{equation}
where the elementary symmetric functions are given as:
\begin{align}
e_0(\lambda_{p,0},\ldots,\lambda_{p,k}) &= 1 \nonumber\\
e_1(\lambda_{p,0},\ldots,\lambda_{p,k}) &= \sum_{0\leq a \leq k}\lambda_{p,a} \nonumber\\
e_2(\lambda_{p,0},\ldots,\lambda_{p,k}) &= \sum_{0\leq a < b \leq k}\lambda_{p,a}\, \lambda_{p,b} \nonumber\\
e_3(\lambda_{p,0},\ldots,\lambda_{p,k}) &= \sum_{0\leq a < b < c \leq k}\lambda_{p,a}\, \lambda_{p,b} \,\lambda_{p,c} \nonumber \\
e_{k+1}(\lambda_{p,0},\ldots,\lambda_{p,k}) &= \lambda_{p,0}\, \lambda_{p,1} \,\ldots \, \lambda_{p,k}
\label{SymmFunc}
\end{align}
The coefficients $A_{p,n}$ and functions $e_{n}$ are related as:
\begin{equation}
A_{p,n} = (-1)^{k+1-n}\, e_{k+1-n} ~.
\end{equation}
The important result of this paper is to show that the coefficients $A_{p,n}$ and hence $e_n$ are always rational numbers, even though the eigenvalues $\lambda_{p,a}$ are irrational.
\begin{mdframed}[style=sid]\textbf{Result.} \emph{The characteristic polynomial of the reduced density matrix $Y_p$ is a monic polynomial of degree $(k+1)$ with rational coefficients.}
\end{mdframed}
In the following subsections, we will show various calculations that support the result.
\subsection{Characteristic polynomial for $T_{2,2}$ (Hopf link)}
The eigenvalues of the reduced density matrix $Y_2$ for the Hopf link are given as:
\begin{equation}
\lambda_{2,0}=\lambda_{2,1} = \lambda_{2,2}=\ldots= \lambda_{2,k} = \frac{1}{k+1} ~.
\end{equation}
The characteristic polynomial of $Y_2$, in variable $x$, will be:
\begin{equation} 
\text{CP}_2(x) = \prod_{a=0}^k(x-\lambda_{2,a}) = \left(x-\frac{1}{k+1}\right)^{k+1} ~.
\end{equation}
The $\text{CP}_2$ can be written as a polynomial in $x$ of degree $(k+1)$ as:
\begin{equation}\text{CP}_2(x) = \sum_{n=0}^{k+1} A_{2,n}\,x^n ~,
\end{equation}
where the coefficients $A_{2,n}$ are given as:
\begin{equation}
\boxed{ A_{2,n} = \frac{(-1)^{k + 1 - n}\,(k+1)!}{(k + 1)^{k + 1-n}\,n!\, (k+1-n)!}
} ~.
\label{A2nCoeff}
\end{equation}
The rationality of the coefficients is obvious here, and the characteristic polynomials for some low values of $k$ are tabulated in table \ref{Tableforp2}. 
\begin{table}[htb]
\centering
\begin{tabular}{|c|l|} \hline  
\rowcolor{gray!50}
$k$ & Characteristic polynomial $\text{CP}_2$ for reduced density matrix $Y_2$ \\ \hline \rule{0pt}{10pt}
0 & $x-1$\\  \hline  \rule{0pt}{15pt}
1 & $x^2-x+\dfrac{1}{4}$ \\[0.2cm]  \hline \rule{0pt}{15pt}
2 & $x^3-x^2+\dfrac{1}{3}x-\dfrac{1}{27}$\\[0.2cm]  \hline \rule{0pt}{15pt}
3 & $x^4-x^3+\dfrac{3}{8}x^2-\dfrac{1}{16}x+\dfrac{1}{256}$ \\[0.2cm] \hline \rule{0pt}{15pt}
4& $x^5-x^4+\dfrac{2}{5}x^3-\dfrac{2}{25}x^2+\dfrac{1}{125}x-\dfrac{1}{3125}$ \\[0.2cm] \hline \rule{0pt}{15pt} 
5 & $x^6-x^5+\dfrac{5}{12}x^4-\dfrac{5}{54}x^3+\dfrac{5}{432}x^2-\dfrac{1}{1296}x+\dfrac{1}{46656}$\\[0.2cm]  \hline 
    \end{tabular}      
    \caption[Characteristic polynomial of RDM for $T_{2,2}$ link for $(1|1)$ bi-partition]{$\text{CP}_2$  polynomials for various values of $k$}
    \label{Tableforp2}
\end{table}
\subsection{Characteristic polynomial for $T_{3,3}$ link}
The eigenvalues of the reduced density matrix $Y_3$ for the $T_{3,3}$ link are given as:
\begin{equation}
\lambda_{3,0}=\frac{(\mathcal{S}_{00})^{-2}}{N_3} \,\,,\,\, \lambda_{3,1} = \frac{(\mathcal{S}_{01})^{-2}}{N_3} \,\,,\,\, \ldots \,\,,\,\, \lambda_{3,k} = \frac{(\mathcal{S}_{0k})^{-2}}{N_3} ~.
\end{equation}
The characteristic polynomial, in variable $x$, will be:
\begin{equation}
\text{CP}_3(x) = \prod_{a=0}^k(x-\lambda_{3,a}) = \prod_{a=0}^k\left(x-\frac{(\mathcal{S}_{0a})^{-2}}{N_3}\right) ~. 
\end{equation}
The characteristic polynomials for some low values of $k$ are tabulated in table \ref{Tableforp3}. 
\begin{table}[htb]
\centering
\begin{tabular}{|c|l|} \hline  
\rowcolor{gray!50}
$k$ & Characteristic polynomial $\text{CP}_3$ for reduced density matrix $Y_3$ \\ \hline \rule{0pt}{10pt}
0 & $x-1$\\  \hline  \rule{0pt}{15pt}
1 & $x^2-x+\dfrac{1}{4}$ \\[0.2cm]  \hline \rule{0pt}{15pt}
2 & $x^3-x^2+\dfrac{8}{25}x-\dfrac{4}{125}$\\[0.2cm]  \hline \rule{0pt}{15pt}
3 & $x^4-x^3+\dfrac{7}{20}x^2-\dfrac{1}{20}x+\dfrac{1}{400}$ \\[0.2cm] \hline \rule{0pt}{15pt}
4& $x^5-x^4+\dfrac{64}{175}x^3-\dfrac{2592}{42875}x^2+\dfrac{6912}{1500625}x-\dfrac{6912}{52521875}$ \\[0.2cm] \hline \rule{0pt}{15pt} 
5 & $x^6-x^5+\dfrac{3}{8}x^4-\dfrac{15}{224}x^3+\dfrac{11}{1792}x^2-\dfrac{1}{3584}x+\dfrac{1}{200704}$\\[0.2cm]  \hline 
    \end{tabular}      
    \caption[Characteristic polynomial of RDM for $T_{3,3}$ link for $(1|2)$ bi-partition]{$\text{CP}_3$ polynomials for various values of $k$}
    \label{Tableforp3}
\end{table}
 Interestingly, we see that the coefficients of the polynomials are rational numbers. Analyzing the pattern of the coefficients, we find that the $\text{CP}_3$ is a monic polynomial in $x$ of degree $(k+1)$ written as:
\begin{equation}\text{CP}_3(x) = \sum_{n=0}^{k+1} A_{3,n}\,x^n ~,
\end{equation}
where the rational coefficients have a close form expression given as:
\begin{equation}
\boxed{ A_{3,n} = \frac{(-3)^{k-n+1} 2^{2 k-2 n+3}\,  \Gamma(2 k+4-n)}{(k+1)^{k+1-n}\, (k+3)^{k+1-n}(k+2)\, \Gamma (n+1)\, \Gamma (2 k+5-2 n)} } ~.
\label{A3nCoeff}
\end{equation}
The above expression makes the rationality of $A_{3,n}$ manifest. For links $T_{p,p}$ with $p \geq 4$, we have the following result:
\begin{mdframed}[style=sid]
\textbf{Result.} \emph{The characteristic polynomial of the reduced density matrix $Y_p$ for $p \geq 4$ can be written as the product of characteristic polynomials of $Y_3$ with appropriate variables. The rationality of $A_{3,n}$ ensures that all other characteristic polynomials will also have rational coefficients.}
\end{mdframed}
To elaborate on the above result, we calculate the characteristic polynomials of $Y_4$ and $Y_5$ and finally give a generic result for $Y_p$.
\subsection{Characteristic polynomial for $T_{4,4}$ link} 
The eigenvalues of the reduced density matrix $Y_4$ for the $T_{4,4}$ link are given as:
\begin{equation}
\lambda_{4,0}=\frac{(\mathcal{S}_{00})^{-4}}{N_4} \,\,,\,\, \lambda_{4,1} = \frac{(\mathcal{S}_{01})^{-4}}{N_4} \,\,,\,\, \ldots \,\,,\,\, \lambda_{4,k} = \frac{(\mathcal{S}_{0k})^{-4}}{N_4} ~.
\end{equation}
We note that these eigenvalues can be written in terms of eigenvalues of $Y_3$ matrix as:
\begin{equation}
\lambda_{4,a} = \frac{N_3^2}{N_4} \lambda_{3,a}^2 ~.
\end{equation}
The characteristic polynomial, in variable $x$, will be:
\begin{align}
\text{CP}_4(x) &= \prod_{a=0}^k(x-\lambda_{4,a}) = \prod_{a=0}^k\left(x-\frac{N_3^2}{N_4} \lambda_{3,a}^2 \right) = \frac{N_3^{2k+2}}{N_4^{k+1}} \prod_{a=0}^k\left(y- \lambda_{3,a}^2 \right) \nonumber \\ 
&= \frac{N_3^{2k+2}}{N_4^{k+1}} \prod_{a=0}^k\left[(y^{1/2}- \lambda_{3,a})(y^{1/2}+ \lambda_{3,a}) \right] \nonumber \\
&= \frac{N_3^{2k+2}}{N_4^{k+1}} (-1)^{k+1}\, \text{CP}_3(y^{1/2})\, \text{CP}_3(-y^{1/2})
\label{CP4asCP3}
\end{align}
where the variable $y$ is:
\begin{equation}
y = \left(\frac{N_4}{N_3^2}\right)x ~.
\end{equation}
Note that it was shown in \cite{Dwivedi:2020rlo} that $N_p$ is an integer. In fact, the functional form of $N_3$ and $N_4$ are given as \cite{Dwivedi:2020rlo}: 
\begin{align}
N_3 = \frac{(k+1) (k+2) (k+3)}{6}   \, ;\,   
N_4 = \frac{(k+1) (k+2)^2 (k+3) \left(k^2+4 k+15\right)}{180} 
\end{align}
Since $\text{CP}_3$ is a polynomial with rational coefficients, this makes $\text{CP}_4$ a polynomial  
\begin{equation}
\text{CP}_4(x) = \sum_{n=0}^{k+1} A_{4,n}\,x^n ~,
\label{CP4}
\end{equation}
with coefficients $A_{4,n}$ as rational numbers. Equating \eqref{CP4} and \eqref{CP4asCP3}, we can write:
\begin{equation}
\sum_{n=0}^{k+1} A_{4,n}\,x^n = \frac{N_3^{2k+2}}{N_4^{k+1}} (-1)^{k+1}\,\sum_{r=0}^{k+1}\sum_{s=0}^{k+1}(-1)^s\,A_{3,r}\,A_{3,s}\,y^{(r+s)/2} ~.
\end{equation}
From this, we can extract $A_{4,n}$ explicitly in terms of $A_{3,n}$ as:  
\begin{equation}
\boxed{A_{4,n} = (-1)^{n+k+1}\,\frac{N_3^{2k+2-2n}}{N_4^{k+1-n}} \left( A_{3,n}^2+\sum_{m=n+1}^{k+1}2(-1)^{m+n}\,A_{3,m}\,A_{3,2n-m} \right) }
\end{equation}
For completeness, we present the characteristic polynomials $\text{CP}_4(x)$ for some low values of $k$ in table \ref{Tableforp4}.
\begin{table}[htb]
\centering
\begin{tabular}{|c|l|} \hline  
\rowcolor{gray!50}
$k$ & Characteristic polynomial $\text{CP}_4$ for reduced density matrix $Y_4$ \\ \hline \rule{0pt}{10pt}
0 & $x-1$\\  \hline  \rule{0pt}{15pt}
1 & $x^2-x+\dfrac{1}{4}$ \\[0.2cm]  \hline \rule{0pt}{15pt}
2 & $x^3-x^2+\dfrac{8}{27}x-\frac{16}{729}$\\[0.2cm]  \hline \rule{0pt}{15pt}
3 & $x^4-x^3+\dfrac{11}{36}x^2-\dfrac{1}{36}x+\dfrac{1}{1296}$ \\[0.2cm] \hline \rule{0pt}{15pt}
4& $x^5-x^4+\dfrac{704}{2303}x^3-\dfrac{1009152}{35611289}x^2+\dfrac{11943936}{11716114081}x-\dfrac{47775744}{3854601532649}$ \\[0.2cm] \hline \rule{0pt}{15pt} 
5 & $x^6-x^5+\dfrac{17}{56}x^4-\dfrac{43}{1568}x^3+\dfrac{13}{12544}x^2-\dfrac{3}{175616}x+\dfrac{1}{9834496}$\\[0.2cm]  \hline 
    \end{tabular}      
    \caption[Characteristic polynomial of RDM for $T_{4,4}$ link for $(1|3)$ bi-partition]{$\text{CP}_4$ polynomials for various values of $k$}
    \label{Tableforp4}
\end{table}
\subsection{Characteristic polynomial for $T_{5,5}$ link} 
The eigenvalues of the reduced density matrix $Y_5$ for the $T_{5,5}$ link are given as:
\begin{equation}
\lambda_{5,0}=\frac{(\mathcal{S}_{00})^{-6}}{N_5} \,\,,\,\, \lambda_{5,1} = \frac{(\mathcal{S}_{01})^{-6}}{N_5} \,\,,\,\, \ldots \,\,,\,\, \lambda_{5,k} = \frac{(\mathcal{S}_{0k})^{-6}}{N_5} ~.
\end{equation}
We note that these eigenvalues can be written in terms of eigenvalues of matrix $Y_3$ as:
\begin{equation}
\lambda_{5,a} = \frac{N_3^3}{N_5} \lambda_{3,a}^3
\end{equation}
Therefore, the characteristic polynomial in variable $x$ will be:  
\begin{align}
\text{CP}_5(x) &= \prod_{a=0}^k(x-\lambda_{5,a}) = \prod_{a=0}^k\left(x-\frac{N_3^3}{N_5} \lambda_{3,a}^3 \right) = \frac{N_3^{3k+3}}{N_5^{k+1}} \prod_{a=0}^k\left(y- \lambda_{3,a}^3 \right) \nonumber \\
& = \frac{N_3^{3k+3}}{N_5^{k+1}} \prod_{a=0}^k\left[(y^{1/3}- \lambda_{3,a})(y^{1/3} - \omega\,\lambda_{3,a}) (y^{1/3}- \omega^2\,\lambda_{3,a}) \right] \nonumber \\
&= \frac{N_3^{3k+3}}{N_5^{k+1}} \omega^{k+1}\omega^{2(k+1)}\, \text{CP}_3(y^{1/3})\,  \text{CP}_3(y^{1/3} \omega^{-1})\, \text{CP}_3(y^{1/3}\omega^{-2}) \nonumber \\
&= \frac{N_3^{3k+3}}{N_5^{k+1}} \, \text{CP}_3(y^{1/3})\,  \text{CP}_3(y^{1/3} \omega^{-1})\, \text{CP}_3(y^{1/3}\omega^{-2})
\label{CP5asCP3}
\end{align}  
where we have:
\begin{equation}
y = \left(\frac{N_5}{N_3^3}\right)x  \quad;\quad \omega = \exp\left(\frac{2\pi i}{3}\right) 
\end{equation}
Since $N_3$ and $N_5$ are integers, the $\text{CP}_5$ will be a polynomial with rational coefficients, i.e.
\begin{equation}
\text{CP}_5(x) = \sum_{n=0}^{k+1} A_{5,n}\,x^n ~,
\end{equation} 
where $A_{5,n}$ are rational coefficients which can be extracted in terms of $A_{3,n}$. For completeness, we tabulate the characteristic polynomials for low values of $k$ in the table \ref{Tableforp5}.
\begin{table}[htb]
\centering
\begin{tabular}{|c|c|} \hline
\rowcolor{gray!50}
$k$ & Characteristic polynomial $\text{CP}_5$ for reduced density matrix $Y_5$ \\ \hline \rule{0pt}{10pt}
0 & $x-1$\\  \hline  \rule{0pt}{15pt}
1 & $x^2-x+\dfrac{1}{4}$ \\[0.2cm]  \hline \rule{0pt}{15pt}
2 & $x^3-x^2+\dfrac{80}{289}x-\dfrac{64}{4913}$\\[0.2cm]  \hline \rule{0pt}{15pt}
3 & $x^4-x^3+\dfrac{11}{40}x^2-\dfrac{1}{80}x+\dfrac{1}{6400}$ \\[0.2cm] \hline \rule{0pt}{15pt}
4& $x^5-x^4+\dfrac{3529216}{13039321}x^3-\dfrac{489037824}{47084988131}x^2+\dfrac{22932357120}{170023892141041}x-\dfrac{330225942528}{613956274521299051}$ \\[0.2cm] \hline \rule{0pt}{15pt} 
5 & $x^6-x^5+\dfrac{309}{1156}x^4-\dfrac{171}{19652}x^3+\dfrac{67}{668168}x^2-\dfrac{5}{11358856}x+\dfrac{1}{1544804416}$\\[0.2cm]  \hline 
    \end{tabular}      
    \caption[Characteristic polynomial of RDM for $T_{5,5}$ link for $(1|4)$ bi-partition]{$\text{CP}_5$ polynomials for various values of $k$}
    \label{Tableforp5}
\end{table}
\subsection{Characteristic polynomial for $T_{p,p}$ link} 
The eigenvalues of the reduced density matrix $Y_p$ for the $T_{p,p}$ link are given as:
\begin{equation}
\lambda_{p,0}=\frac{(\mathcal{S}_{00})^{4-2p}}{N_p} \,\,,\,\, \lambda_{p,1} = \frac{(\mathcal{S}_{01})^{4-2p}}{N_p} \,\,,\,\, \ldots \,\,,\,\, \lambda_{p,k} = \frac{(\mathcal{S}_{0k})^{2p-4}}{N_p} ~.
\end{equation}
These eigenvalues can be written in terms of eigenvalues of the matrix $Y_3$ as:
\begin{equation}
\lambda_{p,a} = \frac{N_3^{p-2}}{N_p} \lambda_{3,a}^{p-2}
\end{equation}
Therefore, the characteristic polynomial in variable $x$ will be:  
\begin{align}
\text{CP}_p(x) &= \prod_{a=0}^k(x-\lambda_{p,a}) = \prod_{a=0}^k\left(x-\frac{N_3^{p-2}}{N_p} \lambda_{3,a}^{p-2} \right) = \frac{N_3^{(p-2)(k+1)}}{N_p^{k+1}} \prod_{a=0}^k\left(y- \lambda_{3,a}^{p-2} \right) \nonumber \\
& = \frac{N_3^{(p-2)(k+1)}}{N_p^{k+1}} \prod_{a=0}^k \prod_{i=0}^{p-3}\left[y^{1/(p-2)}- \theta^i \lambda_{3,a})\right] \nonumber \\
&= \frac{N_3^{(p-2)(k+1)}}{N_p^{k+1}} (-1)^{(p-1)(k+1)} \prod_{i=0}^{p-3} \text{CP}_3(\theta^{-i} y^{1/(p-2)})
\label{CPpasCP3}
\end{align} 
where we have defined
\begin{equation}
y = \left(\frac{N_p}{N_3^{p-2}}\right)x  \quad;\quad  \theta = \exp\left(\frac{2\pi i}{p-2}\right) ~.
\end{equation}  
Since $N_p$ are integers, the equation \eqref{CPpasCP3} ensures that the characteristic polynomial $\text{CP}_p$ will be a polynomial with rational coefficients.
\section{Conclusion} \label{sec3}
In this article, we considered the quantum states $\ket{S^3 \backslash T_{p,p}}$ prepared in SU(2) Chern-Simons theory on the torus link complement. In particular, we have analyzed the reduced density matrix $Y_p$ arising from the $(1|p-1)$ bi-partition of the total Hilbert space. The reduced density matrix plays a key role in quantum entanglement and measurement since it allows for the extraction of all physical quantities related to the reduced degrees of freedom. In particular, our focus was to compute the characteristic polynomials for $Y_p$, which we have denoted as $\text{CP}_p(x)$ where $x$ is some random variable. These characteristic polynomials are monic polynomials of degree $(k+1)$ in variable $x$. We find that the coefficients of these polynomials are always rational numbers. This is an interesting result, given the fact that the eigenvalues themselves are irrational numbers, which is obvious from \eqref{eigen-Tpp}. Note that these coefficients can be written as elementary symmetric functions of the eigenvalues as mentioned in \eqref{SymmFunc}. The rationality of the coefficients for all values of $p$ and $k$ hints at some underlying number-theoretic identity, and it will be interesting and worthwhile to explore this identity further. 

In this work, we have only focused on a particular type of torus link, which is $T_{p,p}$. There are other torus links as well of type $T_{p,q}$ where $p\neq q$. It remains to be seen whether the results of this work are also true for this more general class of torus links. We believe that these results can be systemized and will have a strong implication in number theory. Further, since the roots of these polynomials are precisely the eigenvalues of the reduced density matrix, it will also help generate identities involving various entanglement measures. We leave these open problems as a future endeavor. 

\vspace{0.5in}

\noindent \textbf{Acknowledgment}\\
SD is supported by the “SERB Start-Up Research Grant SRG/2023/001023”.

\bibliographystyle{JHEP}
\bibliography{SU2DM}

\end{document}